# Two-fold symmetry of in-plane magnetoresistance anisotropy in the superconducting states of BiCh$_2$-based LaO$_{0.9}$F$_{0.1}$BiSSe single crystal

Kazuhisa Hoshi[1], Motoi Kimata[2], Yosuke Goto[1], Akira Miura[3], Chikao Moriyoshi[4], Yoshihiro Kuroiwa[4], Masanori Nagao[5], Yoshikazu Mizuguchi[1]*

1. Department of Physics, Tokyo Metropolitan University, 1-1, Minami-osawa, Hachioji 192-0397, Japan
2. Institute for Materials Research, Tohoku University, Sendai 980-8577, Japan
3. Faculty of Engineering, Hokkaido University, Sapporo 060-8628, Japan
4. Graduate School of Advanced Science and Engineering, Hiroshima University, Higashihiroshima, Hiroshima 739-8526, Japan
5. Center for Crystal Science and Technology, University of Yamanashi, Kofu 400-8511, Japan

Corresponding author: Yoshikazu Mizuguchi (mizugu@tmu.ac.jp)

Abstract

Recently, two-fold symmetric in-plane anisotropy of the superconducting properties have been observed in a single crystal of BiCh$_2$-based (Ch: S, Se) layered superconductor LaO$_{0.5}$F$_{0.5}$BiSSe having a tetragonal (four-fold-symmetric) in-plane structure; the phenomena are very similar to those observed in nematic superconductors. To explore the origin of the two-fold symmetric anisotropy in the BiCh$_2$-based system, we have investigated the electron-doping dependence on the anisotropy by examining the in-plane anisotropy of the magnetoresistance in the superconducting states for a single crystal of LaO$_{0.9}$F$_{0.1}$BiSSe under high magnetic fields up to 15 T. We observed a two-fold symmetry of in-plane anisotropy of magnetoresistance for LaO$_{0.9}$F$_{0.1}$BiSSe. The results obtained for LaO$_{0.9}$F$_{0.1}$BiSSe are quite similar to those observed for LaO$_{0.5}$F$_{0.5}$BiSSe, which has a higher electron doping concentration than LaO$_{0.9}$F$_{0.1}$BiSSe. Our present finding suggests that the emergence of the in-plane symmetry breaking in the superconducting state is robust to the carrier concentration in the series of LaO$_{1-x}$F$_x$BiSSe.

## 1. Introduction

The BiCh$_2$-based (Ch: S, Se) superconductor family [1–3], discovered in 2012, has been extensively studied because of the crystal structure similar to the cuprate- and the iron-based high-transition-temperature (high-$T_c$) superconductors [4,5]. Particularly, the crystal structure of the REOBiCh$_2$ (RE: rear earth) systems resemble to that of the iron pnictide system REOFeAs [5]. The typical REOBiS$_2$ system has a layered crystal structure composed of the REO insulating layer and the BiCh$_2$ conducting layer, as shown in Fig. 1(a) [2,3]. The F substitution for the O site generates electron carriers in the BiCh$_2$ layers, which results in the emergence of superconductivity. The highest $T_c$ among the BiCh$_2$-based superconductors is 11 K for LaO$_{0.5}$F$_{0.5}$BiS$_2$ prepared under high pressure [6,7]. The pairing mechanisms of the superconductivity for the BiCh$_2$-based superconductors have not been clarified. Recent theoretical calculations and angle-resolved photoemission spectroscopy (ARPES) have suggested that the unconventional pairing mechanisms are essential for the superconductivity of the BiCh$_2$-based superconductors [8,9]. Moreover, a study of selenium isotope effect for the BiCh$_2$-based superconductor LaO$_{0.6}$F$_{0.4}$BiSSe revealed that the isotope effect exponent ($\alpha_{Se}$) on $T_c$ was close to zero (-0.04 < $\alpha_{Se}$ < +0.04). This result cannot be explained by the electron-phonon mechanism and supports unconventional paring mechanisms [10].

Recently, particularly in the layered superconductors, *nematicity* has been a hot topic due to the possible relation to the unconventional superconducting states. Electronical nematicity has been observed in various unconventional superconductors, such as the cuprate and the iron-based superconductors. Those superconductors show the rotational symmetry breaking while the structural symmetry has been maintained as above the $T_c$ [11,12]. Also, nematic superconductivity has been observed in A$_x$Bi$_2$Se$_3$ (A = Cu, Sr, Nb) [13-18]. The Cu$_x$Bi$_2$Se$_3$ superconductor is known as an odd-parity superconductor, which can be categorized as topological superconductivity. In the superconducting state of A$_x$Bi$_2$Se$_3$, two-fold symmetry of the superconducting gap was observed while the crystal structure has a hexagonal superconducting plane with six-fold symmetry. The unconventional superconducting states in doped Bi$_2$Se$_3$ are related to the strong spin-orbit coupling and multi-orbital characteristics. The BiCh$_2$-based superconductors also possess a strong spin-orbit coupling due to Bi-6p electrons and local inversion symmetry breaking due to the van der Waals gap of the BiCh$_2$ bilayers [19-22]. In addition, a thin film of a related compound LaOSbSe$_2$ was predicted as the Dirac material [23]. On the basis of these similarities to the Cu$_x$Bi$_2$Se$_3$ system, nematic or related exotic phenomena can be expected for the superconducting states of the BiCh$_2$-based compounds. Indeed, two-fold symmetric in-plane anisotropy was recently observed in the superconducting states of LaO$_{0.5}$F$_{0.5}$BiSSe, whose conducting plane has a four-fold tetragonal symmetry [24]. This observation suggests that the BiCh$_2$ superconductors are promising candidates of nematic superconductor. In this study, we show the in-plane anisotropy of magnetoresistance for LaO$_{0.9}$F$_{0.1}$BiSSe with a smaller electron doping concentration than LaO$_{0.5}$F$_{0.5}$BiSSe. As similar to the previous study, the magnetoresistance of the present sample also showed two-fold-symmetric behavior in the superconducting state, suggesting the universal behavior of nematic-superconductivity-like phenomena is in the LaO$_{1-x}$F$_x$BiSSe systems. This study shows that the BiCh$_2$-based compound family will be a useful platform to study the physics and chemistry of nematic superconductivity in layered materials.

## 2. Experimental details

LaO$_{0.9}$F$_{0.1}$BiSSe single crystals were grown by a high-temperature flux method in an evacuated quartz tube. Polycrystalline LaO$_{0.9}$F$_{0.1}$BiSSe was prepared using the solid-state-reaction method using powders of La$_2$O$_3$ (99.9%), La$_2$S$_3$ (99.9%), Bi$_2$O$_3$ (99.999%), and BiF$_3$ (99.9%) and grains of Bi (99.999%) and Se (99.999%) [25]. A mixture of the starting materials was mixed with a nominal ratio of LaO$_{0.9}$F$_{0.1}$BiSSe, pressed into a pellet and annealed at 700 ºC for 20 h in an evacuated quartz tube. The polycrystalline LaO$_{0.9}$F$_{0.1}$BiSSe (0.62 g) were mixed with CsCl flux (2.2 g), and the mixture was sealed into an evacuated quartz tube. The tube was heated at 900 ºC for 12 h, slowly cooled to 645 ºC with a rate of 1.0 ºC /h, and furnace-cooled to room temperature. After furnace cooling, the quartz tube was opened under air atmosphere, and the product was filtered and washed with pure water. The chemical composition of the obtained crystal was investigated using energy-dispersive X-ray spectroscopy (EDX) spectroscopy. The average compositional ratio of the constituent elements (except for O and F) was estimated to be La : Bi : S : Se = 1 : 0.99 : 0.93 : 1.0, which was normalized by the La value. The analyzed atomic ratio is almost consistent with the nominal composition LaO$_{0.9}$F$_{0.1}$BiSSe. Considering the typical detection error in the EDX analysis, we regard the composition of the examined crystal as LaO$_{0.9}$F$_{0.1}$BiSSe.

The single crystals were ground with quartz powders to get homogeneous powders for the SXRD experiment [26]. The synchrotron powder X-ray diffraction (SXRD) was performed at the beamline BL02B2 SPring-8 at a wavelength of 0.495274 Å (proposal No. 2018B1246). The SXRD experiments were performed with a sample rotator system at room temperature; the diffraction data were collected using a high-resolution one-dimensional semiconductor detector (multiple MYTHEN system [27]) with a step size of $2\theta = 0.006$º. The crystal structure parameters were refined using the Rietveld method with the RIETAN-EP software [28]. The schematic image of the crystal structure refined by the Rietveld refinement was depicted using the VESTA software [29].

The resistive anisotropy was investigated under magnetic fields up to 15 T using a superconducting magnet at the Institute for Materials Research (IMR) of Tohoku University To precisely investigate the anisotropy, a $^3$He probe equipped with an accurate two-axes rotator system was used. The magnetic field perpendicular to the $c$-axis was applied. The electrodes were fabricated using Au wires and Ag pastes.

3. **Results and discussion**

Figure 1(b) shows the SXRD pattern for the powdered sample of LaO$_{0.9}$F$_{0.1}$BiSSe. The SXRD pattern was refined using the tetragonal ($P4/nmm$) structural model. The refined lattice parameters are $a$ = 4.10849(9) Å and $c$ = 13.6851(4) Å. The lattice parameters $a$ and $c$ are slightly smaller than those obtained with a polycrystalline sample with $x$ = 0.1 in Ref. 25, and the lattice constant $c$ for the crystal was comparable to that observed for $x$ = 0.3. Since the lattice constant sometimes differs between single and polycrystalline samples in BiCh$_2$-based superconductors, we use the analysis result on the lattice constant for the confirmation that electron carrier has been doped in the crystal. As mentioned later, from the estimation of $T_c$, we could assume that the carrier doping amount is close to the starting nominal value of $x$ = 0.1. In the Rietveld refinement, we assumed that the in-plane Ch1 site [see Fig. 1(a)] was fully occupied by Se, and the Ch2 site was fully occupied by S, on the basis of the EDX analysis result and previous structural analysis [25]. The obtained reliable factor $R_{wp}$ was $R_{wp}$ = 11.5%. In fact, the 00$l$ peaks and others related to $c$-axis direction are broadened and have a shoulder, which should be resulting in a slightly high $R_{wp}$. This may be due to the strain introduced during the sample preparation by grinding with quartz powders and the sticky nature of the crystals. However, the higher angle fitting with the tetragonal model (see the inset of Fig. 1) is quite nice. In addition, we did not obtain better $R_{wp}$ with monoclinic model. Although the LaO$_{1-x}$F$_x$BiSSe phase undergoes a structural transition from tetragonal to monoclinic ($P2_1/m$), our recent study suggested that the transition is suppressed by 3% substitution of O by F [30]. Therefore, the crystal structure of the present crystal with nominal $x$ = 0.1 can be regarded as tetragonal with four-fold symmetry in the conducting plane down to low temperature near $T_c$.

Figure 2 shows the temperature dependence of electrical resistance for LaO$_{0.9}$F$_{0.1}$BiSSe single crystal measured with current along the $c$-axis. The $T_c^{onset}$ is 3.4 K, and the $T_c^{zero}$ is 2.9 K. These $T_c$ values are comparable to those observed for polycrystalline sanples with $x$ = 0.1. Figure 3(a) shows the magnetic field dependence of resistance where the magnetic field perpendicular to the $c$-axis was applied. The superconducting states are suppressed with increasing magnetic field. Figure 3(b) shows the temperature dependence of the in-plane upper critical field $H_{c2}$, estimated from the midpoint $T_c$ ($T_c^{mid}$) in the superconducting transition, for LaO$_{0.9}$F$_{0.1}$BiSSe. From the liner extrapolation of the data reaches 20 T at 0 K. The $\mu_0H_{c2}(0)$ estimated from Werthamer-Helfand-Hohemberg model [31] was 13.8 T. The high $H_{c2}$ with $H$ // ab is consistent with previous reports for BiCh$_2$-based superconductors [3,22].

Figure 4(a) shows a schematic image of the terminal configuration for the in-plane anisotropy measurement. To investigate the in-plane anisotropy in the superconducting states, the crystal was rotated using two rotation angles. $\theta$ and $\phi$ are defined as shown in Fig 4(b). $\theta$ is defined as the formed angle from the *c*-axis to *ab*-plane and $\phi$ is defined as the formed angle from the *a*-axis to the *b*-axis. The angle dependences of resistance were investigated for the superconducting states in between the onset and the zero-resistance states by tuning the temperature and magnetic field based on the obtained field-temperature phase diagram [Fig. 3(b)].

Figure 4(c) shows the $\theta$ angle dependence of the electronical resistance at $\phi = 180°$, $\mu_0 H = 4$ T, and $T = 2$ K. The $R_{min}$ is defined as the minimum electronical resistance, which was at $\theta = 90°$ in the angle scan. Figure 4(d) shows the $\phi$ angle dependence of $R_{min}$ where the in-plane anisotropy of the magnetoresistance in the superconducting states (or upper critical field) can be investigated from the $\phi$ scan of $R_{min}$. Although the crystal possessed a four-fold symmetry in its in-plane structure as discussed above, an oscillation of $R_{min}$ with a period of about 180° in Fig. 4(d). Furthermore, the plotted data was well fitted by the function of $A\sin(2\phi + \alpha) + B\sin(4\phi + \beta) + C$. The estimated amplitude constants $A$, $B$, and $C$ were 0.034(2), -0.0076(24), and 0.1198(17). Since the constant $A$ related to the two-fold symmetry oscillation is clearly larger than $B$ related to the four-fold symmetry, the appearance of the two-fold symmetry in the in-plane anisotropy of the upper critical field should be essential.

As introduced in the introduction, similar two-fold-symmetric in-plane anisotropy of magnetoresistance in the superconducting states has been observed for $LaO_{0.5}F_{0.5}BiSSe$. Since the present experiments revealed that $LaO_{0.9}F_{0.1}BiSSe$ also exhibits the two-fold symmetry in the superconducting states, we conclude that the carrier concentration is not an essential parameter for the condition of the appearance of the phenomena in the $LaO_{1-x}F_xBiSSe$ system. This fact surprised us because the Fermi surface topology is largely different between electron doping concentrations of $x = 0.1$ and $x = 0.5$ in the $BiCh_2$-based compounds [32]. The phenomena similar to nematic superconductivity states in doped $Bi_2Se_3$ systems will motivate further experiments in $LaO_{1-x}F_xBiSSe$ and related systems. Although there are no theoretical studies predicting the emergence of nematic superconductivity in the $BiCh_2$-based compounds, a theoretical study predicted the possible topological superconductivity in $BiCh_2$-based systems [33]. Since the pairing mechanisms of the $BiCh_2$-based superconductors have not been concluded [32], further theoretical and experimental investigations are needed, and the present result on the in-plane anisotropy of the superconducting states in $LaO_{1-x}F_xBiSSe$ should be one of the key information for the goal.

## 4. Conclusion

We have investigated the transport properties of a single crystal of $BiCh_2$-based superconductor $LaO_{0.9}F_{0.1}BiSSe$ under high magnetic fields up to 15 T. From the *c*-axis electrical resistance (measured with a current along the *c*-axis), the upper critical field was determined. Also, the in-plane anisotropy of the electrical resistance was investigated using a $^3$He probe equipped with a two-axes rotator system to investigate the in-plane anisotropy of magnetoresistance. From the in-plane anisotropy measurements, we observed two-fold symmetry of magnetoresistance in the superconducting states within the *ab* plane of $LaO_{0.9}F_{0.1}BiSSe$. Since the crystal possessed a tetragonal square plane with a tetragonal (four-fold) in-plane symmetry, the appearance of two-fold symmetry indicates the rotational symmetry breaking in the superconducting states. The phenomena are very similar to those observed for $LaO_{0.5}F_{0.5}BiSSe$ with a higher electron doping concentration. Therefore, we conclude that the carrier doping concentration, which affects the Fermi surface topology, is not an essential parameter for the emergence of the nematic-superconductivity-like phenomena in $LaO_{1-x}F_{x}BiSSe$. We hope that the results shown here are useful for further investigation on superconductivity pairing mechanisms of the $BiCh_2$-based compounds and related studies on nematic superconductivity in layered systems.

**Acknowledgement**

The authors thank I. Tanaka, O. Miura, M. Kishimoto, R. Omura, and T. D. Matsuda for experimental support and fruitful discussion. This work was partly supported by Collaborative Research with IMR, Tohoku Univ. (proposal number: 17H0074) and Grants-in-Aid for Scientific Research (Nos. 15H05886, 16H04493, 18KK0076, and 19K15291) and the Advanced Research Program under the Human Resources Funds of Tokyo (Grant Number: H31-1).

Reference

[1] Y. Mizuguchi, H. Fujihisa, Y. Gotoh, K. Suzuki, H. Usui, K. Kuroki, S. Demura, Y. Takano, H. Izawa, O. Miura, *Phys. Rev. B* **86**, 220510 (2012).

[2] Y. Mizuguchi, S. Demura, K. Deguchi, Y. Takano, H. Fujihisa, Y. Gotoh, H. Izawa, O. Miura, *J. Phys. Soc. Jpn.* **81**, 114725 (2012).

[3] Y. Mizuguchi, *J. Phys. Soc. Jpn.* **88**, 041001 (2019).

[4] J. G. Bednorz, K. A. Müller, *Z. Phys. B-Condens. Matter* **64**, 189 (1986).

[5] Y. Kamihara, T. Watanabe, M. Hirano, H. Hosono, *J. Am. Chem. Soc.* **130**, 3296 (2008).

[6] K. Deguchi, Y. Mizuguchi, S. Demura, H. Hara, T. Watanabe, S. J. Denholme, M. Fujioka, H. Okazaki, T. Ozaki, H. Takeya, T. Yamaguchi, O. Miura, Y. Takano, *EPL* **101**, 17004 (2013).

[7] Y. Mizuguchi, T. Hiroi, J. Kajitani, H. Takatsu, H. Kadowaki, O. Miura, *J. Phys. Soc. Jpn.* **83**, 053704 (2014).

[8] C. Morice, R. Akashi, T. Koretsune, S. S. Saxena, R. Arita, *Phys. Rev. B* **95**, 180505 (2017).

[9] Y. Ota, K. Okazaki, H. Q. Yamamoto, T. Yamamoto, S. Watanabe, C. Chen, M. Nagao, S. Watauchi, I. Tanaka, Y. Takano, S. Shin, *Phys. Rev. Lett.* **118**, 167002 (2017).

[10] K. Hoshi, Y. Goto, Y. Mizuguchi, *Phys. Rev. B* **97**, 094509 (2018).

[11] J. Wu, A. T. Bollinger, X. He, I. Božović, *Nature* **547**, 432 (2017).

[12] S. Kasahara, H. J. Shi, K. Hashimoto, S. Tonegawa, Y. Mizukami, T. Shibauchi, K. Sugimoto, T. Fukuda, T. Terashima, Andriy H. Nevidomskyy, Y. Matsuda, *Nature* **486**, 382 (2012).

[13] Y. S. Hor, A. J. Williams, J. G. Checkelsky, P. Roushan, J. Seo, Q. Xu, H. W. Zandbergen, A. Yazdani, N. P. Ong, R. J. Cava, *Phys. Rev. Lett.* **104**, 057001 (2010).

[14] S. Yonezawa, K. Tajiri, S. Nakata, Y. Nagai, Z. Wang, K. Segawa, Y. Ando, Y. Maeno, *Nature Phys.* **13**, 123 (2017).

[15] K. Matano, M. Kriener, K. Segawa, Y. Ando, G. Zheng, *Nat. Phys.* **12**, 852 (2016).

[16] G. Du, Y. F. Li, J. Schneeloch, R. D. Zhong, G. D. Gu, H. Yang, H. Lin, H. H. Wen, *Sci. China-Phys. Mech. Astron.* **60**, 037411 (2017).

[17] Y. Pan, A. M. Nikitin, G. K. Araizi, Y. K. Huang, Y. Matsushita, T. Naka, A. de Visser, Sci. Rep. 6, 28632 (2016).

[18] S. Yonezawa, Condens. Matter 4, 2 (2019).

[19] Y. Fuseya, M. Ogata, H. Fukuyama, *J. Phys. Soc. Jpn.* **84**, 012001 (2015).

[20] K. Terashima, J. Sonoyama, T. Wakita, M. Sunagawa, K. Ono, H. Kumigashira, T. Muro, M. Nagao, S. Watauchi, I. Tanaka, H. Okazaki, Y. Takano, O. Miura, Y. Mizuguchi, H. Usui, K. Suzuki, K. Kuroki, Y. Muraoka, T. Yokoya, *Phys. Rev. B* **90**, 220512 (2014).

[21] Y. Ma, Y. Dai, N. Yin, Tao. Jing, B. Huang, *J. Mater. Chem. C* **2**, 8539 (2014).

[22] Y. C. Chan, K. Y. Yip, Y. W. Cheung, Y. T. Chan, Q. Niu, J. Kajitani, R. Higashinaka, T. D. Matsuda, Y. Yanase, Y. Aoki, K. T. Lai, Swee K. Goh, *Phys. Rev. B* **97**, 104509 (2018).

[23] X. Y. Dong, J. F. Wang, R. X. Zhang, W. H. Duan, B. F. Zhu, J. O. Sofo, C. X. Liu, *Nat. Commun.* **6**, 8517 (2015).

[24] K. Hoshi, M. Kimata, Y. Goto, T. D. Matsuda, Y. Mizuguchi, *J. Phys. Soc. Jpn.* **88**, 033704 (2019).

[25] K. Nagasaka, A. Nishida, R. Jha, J. Kajitani, O. Miura, R. Higashinaka, T. D. Matsuda, Y. Aoki, A. Miura, C. Moriyoshi, Y. Kuroiwa, H. Usui, K. Kuroki, Y. Mizuguchi, *J. Phys. Soc. Jpn.* **86**, 074701 (2017).

[26] A. Miura, M. Nagao, Y. Goto, Y. Mizuguchi, T. D. Matsuda, Y. Aoki, C. Moriyoshi, Y. Kuroiwa, Y. Takano, S. Watauchi, I. Tanaka, N. C. Rosero-Navarro, K. Tadanaga, *Inorg. Chem.* **57**, 5364 (2018).

[27] S. Kawaguchi, M. Takemoto, K. Osaka, E. Nishibori, C. Moriyoshi, Y. Kubota, Y. Kuroiwa, K. Sugimoto, *Rev. Sci. Instr.* **88**, 085111 (2017).

[28] F. Izumi and K. Momma, *Solid State Phenom.* **130**, 15 (2007).

[29] K. Momma and F. Izumi, *J. Appl. Crystallogr.* **41**, 653 (2008).

[30] K. Hoshi, S. Sakuragi, T. Yajima, Y. Goto, A. Miura, C. Moriyoshi, Y. Kuroiwa, Y. Mizuguchi, arXiv: 2006.07798.

[31] N. R. Werthamer, E. Helfand, P. C. Hohemberg, *Phys. Rev.* **147**, 295-302 (1966).

[32] K. Suzuki, H. Usui, K. Kuroki, T. Nomoto, K. Hattori, H. Ikeda, *J. Phys. Soc. Jpn.* **88**, 041008 (2019).

[33] Y. Yang, W. S. Wang, Y. Y. Xiang, Z. Z. Li, Q. H. Wang, *Phys. Rev. B* **88**, 094519 (2013).

**Figures**

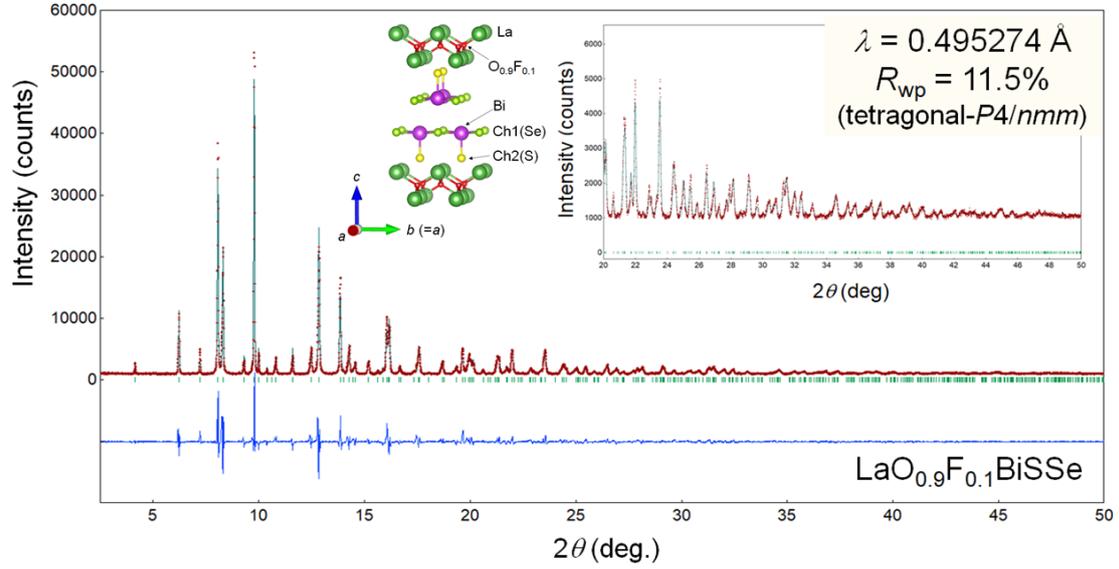

Fig. 1. Synchrotron powder XRD pattern and Rietveld refinement result for the ground $LaO_{0.9}F_{0.1}BiSSe$ crystals. The inset figures show the expanded view at higher angles and a schematic image of crystal structure of $LaO_{0.9}F_{0.1}BiSSe$.

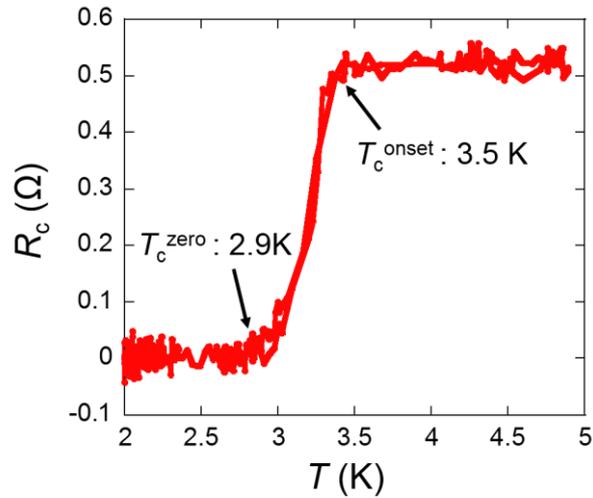

Fig. 2. Temperature dependence of electrical resistance (measured with a current along the $c$-axis) of $LaO_{0.9}F_{0.1}BiSSe$ single crystal from 2 to 5 K at 0 T.

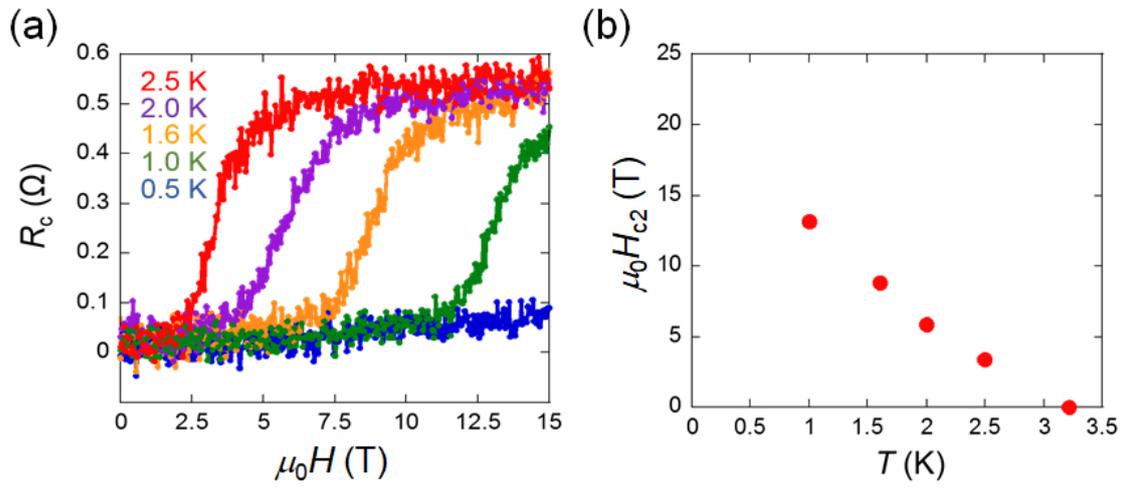

Fig. 3. (a) In-plane field dependence of electronical resistance at 2.5 K, 2.0 K, 1.6 K, 1.0 K and 0.6 K, respectively. (b) The temperature dependence of $H_{c2}$.

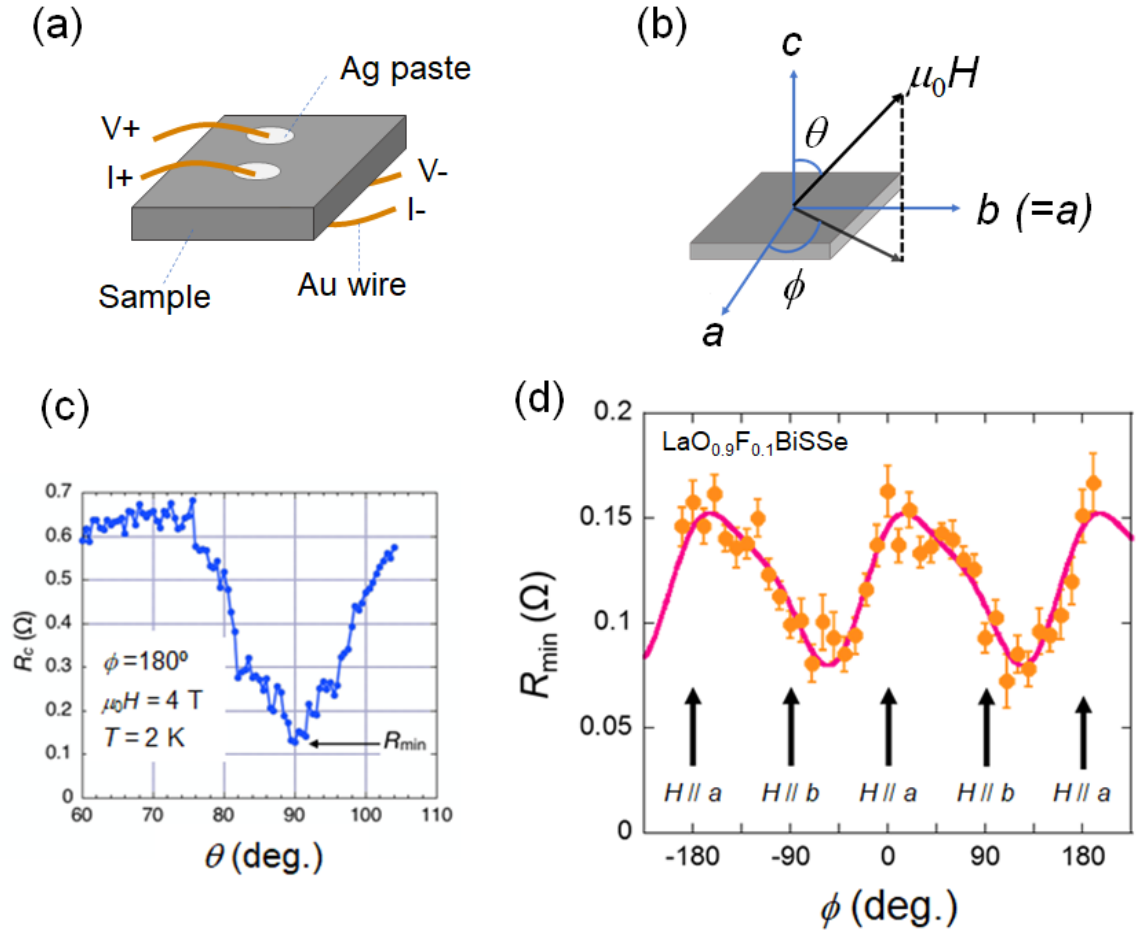

Fig. 4. (a,b) Schematic images of the anisotropy measurement; the terminal configuration and the definition of angles used in the anisotropy analyses. (c) $\theta$ angle dependence of the interface electronical resistance at $\phi = 180°$, $\mu_0 H = 4$ T, and $T = 2$ K. (d) $\phi$ angle dependence of the $R_{min}$. The plotted data was fitted by the function of $A\sin(2\phi + \alpha) + B\sin(4\phi + \beta) + C$.